\documentclass{article}
\usepackage{mdframed}
\usepackage{spconf,amsmath,graphicx}
\usepackage{algorithm}
\usepackage{algorithmic}
\usepackage{graphicx}
\usepackage{caption}
\usepackage{subcaption}
\usepackage{cite}
\usepackage{amsfonts}
\usepackage{fancyhdr}
\usepackage{amsmath,amssymb,amsthm}%
\usepackage{listings}%
\usepackage{arydshln}
\usepackage{subfig}%
\usepackage{algorithmic}%
\title{Sparsity-Exploiting Anchor Placement for Localization \\ in Sensor Networks}
\name{Sundeep Prabhakar Chepuri, Geert Leus, and Alle-Jan van der Veen\thanks{This work was supported in part by STW under the FASTCOM project (10551) and in part by NWO-STW under the VICI program (10382). }}
\address{Faculty of Electrical Engineering, Mathematics, and Computer Science (EEMCS)\\
Delft University of Technology (TU Delft), The Netherlands\\
Email:~\{s.p.chepuri; g.j.t.leus; a.j.vanderveen\}@tudelft.nl.}
\begin{document}
%
%





\maketitle
\begin{abstract}
We consider the anchor placement problem in localization based on one-way ranging, in which either the sensor or the anchors send the ranging signals. 
The number of anchors deployed over a geographical area is generally sparse, and we show that the anchor placement can be formulated as the design of a sparse selection vector. Interestingly, the case in which the anchors send the ranging signals, results in a joint ranging energy optimization and anchor placement problem. We make abstraction of the localization algorithm and instead use the Cram\'er-Rao lower bound (CRB) as the performance constraint. The anchor placement problem is formulated as an elegant convex optimization problem which can be solved efficiently.
\end{abstract}
\begin{keywords}
Anchor placement, ranging energy optimization, localization, sparsity, convex optimization.
\end{keywords}
\vspace*{-2mm}
\section{Introduction}
\vspace*{-1mm}
Localization is an important and extensively studied topic in wireless sensor networks (WSNs)~\cite{Gusta05SPM}. 
Localization can be performed using a plethora of algorithms~\cite{Gusta05SPM} (and references therein), which exploit inter-node measurements like time-of-arrival (TOA), time-difference-of-arrival (TDOA), angle-of-arrival (AOA), or received signal strength (RSS).

The performance of any location estimator depends not only on the algorithm  but also on the placement of the anchors (nodes with known locations). Anchor placement is a key challenge in localization system design, as certain anchor positions not only deteriorate the performance but also result in ambiguities or identifiability issues. In~\cite{optanchor1}, the effect of anchor placement is studied using the geometric dilution of precision (GDOP). The idea of GDOP is borrowed from the global positioning system (GPS) and it is obtained from the Cram\'er-Rao lower bound (CRB) with simplifying assumptions on the noise model, i.e., equal variances on all the range estimates. This assumption is valid for GPS due to the approximately equal distances between the anchors (the satellites) and the nodes, but it does not translate well to WSNs where the noise variance is proportional to the distance (typically different) between the anchors and the sensor.  
In~\cite{Koengeo}, the effect of the anchor positions has been studied by empirically identifying the ambiguity issues. However, they do not provide any algorithm for anchor placement. 

The anchor placement problem can be interpreted as the problem where we divide a specific anchor area $\mathcal{A}$ into $M$ grid points and select the positions of the $K$ anchors as the best $K$ grid points out of $M$ grid points, where $K \ll M$. Here, the $K$ selected anchors are deemed the best, if they guarantee a certain minimal accuracy on the location estimates within a specific sensor area $\mathcal{S}$. In practice, $K$ is not known, and this makes anchor selection a combinatorial problem involving an exhaustive search over all the $2^M$ possible anchor positions, and the  computation over $\mathcal{S}$ is even more cumbersome. 

In this paper, we consider TOA-based ranging, however, we do not restrict ourselves to a particular localization algorithm. Instead, we use the CRB as a performance constraint.
The anchor placement problem is studied for the following two cases of TOA-based one-way ranging: a) {\it the anchors send the ranging signals} (OW-A) and b) {\it the sensor sends the ranging signals} (OW-S).
The anchor placement problem is formulated as the design of a sparse selection vector. For the OW-A case, the sparse solution yields the ranging energies\footnote{More details on ranging energy optimization can be found in~\cite{Tao09TSP}, however, the design of optimal anchor positions is not considered in~\cite{Tao09TSP}.} that the anchors should adopt leading to a solution for the {\it joint ranging energy optimization and anchor placement} problem, and for the~OW-S case the sparse solution yields the optimal anchor positions. Using the CRB as a performance constraint, we can formulate the anchor placement problem as a semidefinite programming (SDP) problem.
The major contribution of this paper is the novel framework for anchor placement in WSNs exploiting the inherent sparse nature of the spatially distributed anchors. 

\vspace*{-3mm}
\section{System model and preliminaries} \label{sec:model}
We consider a two-dimensional network with one sensor located in the sensor area $\mathcal{S}$ and $M$ possible anchors located at the $M$ grid points of the anchor area $\mathcal{A}$. Let the coordinates of the sensor and the $m$th anchor be denoted by the $2 \times 1$ vectors ${\bf s}$ and ${\bf a}_m$, respectively, where $\bf{s}$ is assumed to be unknown but known to be within $\mathcal{S}$. We further assume that all the nodes are time-synchronized (using ~\cite{ChepuriSPL} for instance).
Let the pairwise distance between the sensor and the $m$th anchor be denoted by $d({\bf a}_m,{\bf s}) = {\|{\bf a}_m - {\bf s}\|}_2$. In practice, the pairwise distances are obtained by ranging and they are generally noisy. For the sake of simplicity, we consider a TOA-based one-way ranging.

%
\vspace*{-4mm}
\subsection{Anchors send the ranging signals: OW-A}
The ranging signals sent by the anchors are scheduled such that they can be separated at the sensors. Let the $m$th anchor broadcast a ranging signal $\sqrt{e_m}s_a(t)$ of energy $e_m$ at time $T_{m,s}$, and upon reception at the sensor, the TOA $\hat{T}_{s,m}$ is estimated. The range estimate of the sensor to the $m$th anchor is then given by 
\begin{equation}
\label{ eq:TOF}
\begin{aligned}
\hat{d}_{a}({\bf a}_m,{\bf s})  = c (T_{m,s} -\hat{T}_{s,m})
\end{aligned}
\end{equation} where $c$ is the propagation speed of a wave in the medium. 
Under the assumption that a line of sight (LOS) channel exists between the sensor and the $m$th anchor, it is motivated in~\cite{Tao09TSP,yiyincrb} that the range estimate $\hat{d}_{a}({\bf a}_m,{\bf s})$ is Gaussian distributed as $\mathcal{N}(d({\bf a}_m,{\bf s}), \sigma_{m,s}^2)$ with variance $\sigma_{m,s}^2 = e_m^{-1} \frac{\rho_a}{\gamma^2}$. Here, $\rho_{a} = \frac{c^2N_s/2}{\bar{F_a^2}}$ with $\bar{{F}_a^2} = \frac{\int_{-\infty}^{\infty} (2\pi F)^2|S_a(F)|^2 dF}{\int_{-\infty}^{\infty}|S_a(F)|^2 dF}$ the mean square bandwidth of the ranging signal ($S_a(F)$ being the Fourier transform of $s_a(t)$), and  with $N_s/2$ the two-sided power spectral density (PSD) of the additive white Gaussian noise (AWGN) at the sensor. Further, the signal suffers an attenuation of $\gamma^2 = \alpha d({\bf a}_m,{\bf s})^{-\beta}$, with $\alpha$ and $\beta$ the path gain at $1 \, \mathrm{m}$ and path-loss coefficient, respectively.
\vspace*{-4mm}
\subsection{Sensor sends the ranging signal: OW-S}
In the OW-S case, the sensor broadcasts a ranging signal $\sqrt{e_s}s(t)$ of energy $e_s$ at time $T_{s,m}$, and upon reception at the $m$th anchor, the TOA $\hat{T}_{m,s}$ is estimated. As earlier, the range estimate of the sensor to the $m$th anchor is given by
\begin{equation}
\label{eq:TOF2}
\begin{aligned}
\hat{d}_{s}({\bf a}_m,{\bf s})  = c (T_{s,m} -\hat{T}_{m,s})
\end{aligned}
\end{equation}
which is again assumed to be Gaussian distributed with variance $\sigma_{s,m}^2 =e_s^{-1} \frac{\rho_s}{\gamma^2}$,
where $\rho_{s} = \frac{c^2N_a/2}{\bar{F^2}}$ with $\bar{{F}^2} = \frac{\int_{-\infty}^{\infty} (2\pi F)^2|S(F)|^2 dF}{\int_{-\infty}^{\infty}|S(F)|^2 dF}$ the mean square bandwidth of the ranging signal ($S(F)$ being the Fourier transform of $s(t)$), and with $N_a/2$ the two-sided PSD of the AWGN at the anchor.
\vspace*{-1mm}
\section{Performance measure}
\vspace*{-2mm}
We make abstraction of the localization algorithm, however, we assume that the TOA estimates are unbiased and achieve the CRB asymptotically. Even though this assumption is too optimistic for a practical system, the CRB has a very attractive mathematical structure resulting in a selection problem that can be efficiently solved using convex optimization techniques. Moreover, the CRB optimal anchor positions can improve the performance of any practical localization algorithm.  

The CRB for OW-A and OW-S can be derived based on the CRB for TOA-based two-way ranging~\cite{Tao09TSP,yiyincrb}.
\vspace*{-3mm} 
\subsection{CRB for OW-A} 
For OW-A, the CRB of ${\bf s}$ is denoted by ${\bf C}_{a} \in \mathbb{R}^{2 \times 2}$, and is computed as follows  
\begin{eqnarray}
{\bf C}_{a}^{-1} = {\bf F}_{a}({\bf e},{\bf s}) = \sum_{m=1}^{M} e_m {\bf F}_{a,m}({\bf s})
\label{eq:crlb1}
\end{eqnarray}
where ${\bf F}_{a,m}({\bf s}) = \alpha  \rho_{a}^{-1}d({\bf a}_m, {\bf s})^{-\beta-2} ({\bf s}-{\bf a}_m)({\bf s}-{\bf a}_m)^T$ and ${\bf F}_a({\bf e},{\bf s})$ is the Fisher information matrix (FIM). 
Here, the vector ${\bf e} = [e_1,e_2,\ldots,e_M]^T$ is the {\it joint selection and anchor ranging energy} vector that has to be designed, where a non-zero entry of ${\bf e}$ not only indicates that the anchor position is selected but also represents the ranging signal energy that the selected anchor should adopt.
\vspace*{-3mm}
\subsection{CRB for OW-S} 
Similarly for OW-S, the CRB of ${\bf s}$ is denoted by ${\bf C}_{s} \in \mathbb{R}^{2 \times 2}$, and is computed as follows  
\begin{eqnarray}
{\bf C}_{s}^{-1} = {\bf F}_{s}({\bf w},{\bf s}) = \sum_{m=1}^{M} {w}_m{\bf F}_{s,m}({\bf s})
\label{eq:crlb2}
\end{eqnarray}
where ${\bf F}_{s,m}({\bf s}) = e_s \alpha  \rho_{a}^{-1}d({\bf a}_m, {\bf s})^{-\beta-2}  ({\bf s}-{\bf a}_m)({\bf s}-{\bf a}_m)^T$ and
${\bf F}_{s}({\bf w},{\bf s})$ is the FIM. Here, ${\bf w} = [w_1,\ldots,w_M]^T \in \{0,1\}^M$ is the {\it selection} vector to be designed, where $w_m = 1(0)$ indicates that the related anchor is (not) selected.
\vspace*{-4mm}
\subsection{Identifiability and ambiguity}
The FIM is singular when both the anchors and the sensor are collinear for which the CRB will be infinity. Hence, a CRB-optimal anchor placement would avoid the CRB being infinity. However, when only the anchors (even if $M \ge 3$) are collinear (excluding the sensor) a ``mirror" ambiguity is obtained, and this effect cannot be seen with the FIM, and surprisingly the FIM will be non-singular.
Hence, a (local) CRB-optimal anchor placement can still result in such ambiguities. Such solutions can be avoided with additional constraints or prior information on the parameters, e.g., by a constraint on the sensor area,  such as an orthant or a half plane. 

A constrained CRB generally gives a lower bound for parameters with such deterministic constraints. A constrained CRB is derived from the unconstrained CRB and a non-redundant constraint set. However, for orthant or half plane constraints, it is shown in~\cite{ConstrainedCRB} that the unconstrained CRB is the same as the constrained CRB, and does not yield a lower bound. Hence, throughout this paper, we use the unconstrained CRB in~\eqref{eq:crlb1}-\eqref{eq:crlb2} as a performance measure and we will select our $\mathcal{S}$ carefully.   
\vspace*{-2mm}
\subsection{Constraint for accurate positioning}
Using the definition of accurate positioning from the Federal communication commission (FCC)~\cite{Gusta05SPM}, for every ${\bf s}$ within $\mathcal{S}$ we constrain the localization error ${\boldsymbol\xi} = \hat{\bf {s}} -{\bf {s}}$ to be within an origin-centered circle of radius $R_e$ with a probability higher than $P_e$, i.e., $\forall {\bf s} \in \mathcal{S},\mathrm{Pr}(\|\boldsymbol\xi\|_2 \leq R_e) \geq P_e$. The values of $R_e$ and $P_e$ define the accuracy required and are assumed to be known. 

Sufficient conditions to satisfy this accuracy requirement for OW-A and OW-S are $\lambda_{min}({\bf e},{\bf s}) \geq \lambda$ and $\lambda_{min}({\bf w},{\bf s}) \geq \lambda$, respectively~\cite{Gusta05SPM,Tao09TSP}. Here, $\lambda_{min}({\bf e},{\bf s})$ and $\lambda_{min}({\bf w},{\bf s})$ are the smallest eigenvalues of the matrices ${\bf F}_{a}$ and ${\bf F}_{s}$, respectively, and the threshold $\lambda$ for a Gaussian distributed location estimate is given by $\lambda = \frac{2}{R_e^2} \ln(\frac{1}{1-P_e})$ and for an unknown distribution by $\lambda = \frac{2}{R_e^2} (\frac{1}{1-P_e})$~\cite{Gusta05SPM,Tao09TSP}.

More specifically, we lower bound the eigenvalues of the FIMs ${\bf F}_{a}({\bf e} ,{\bf s})$ and ${\bf F}_{s}( {\bf w} ,{\bf s})$, so that the related CRB is upper bounded. The inequality constraints $\lambda_{min}( {\bf e} ,{\bf s}) \geq \lambda$ and $\lambda_{min}( {\bf w} ,{\bf s}) \geq \lambda$ are equivalent to the following linear matrix inequalities (LMIs): $\sum_{m=1}^{M} e_m {\bf F}_{a,m}({\bf s}) \succeq \lambda {\bf I}_2$ and $\sum_{m=1}^{M} w_m {\bf F}_{s,m}({\bf s}) \succeq \lambda {\bf I}_2$, respectively, where
${\bf A} \succeq {\bf B}$ means that ${\bf A} - {\bf B}$ is a positive semidefinite matrix, and ${\bf I}_2$ is a $2 \times 2$ identity matrix. It is well known that the solution set of ${\bf e}$ and ${\bf w}$ satisfying these respective LMIs is convex~\cite{Boyd}.
\begin{figure*}[!th]
        \centering
        \begin{subfigure}[t]{0.33\textwidth}
                \centering
                \includegraphics[width=\columnwidth,height=1.75in]{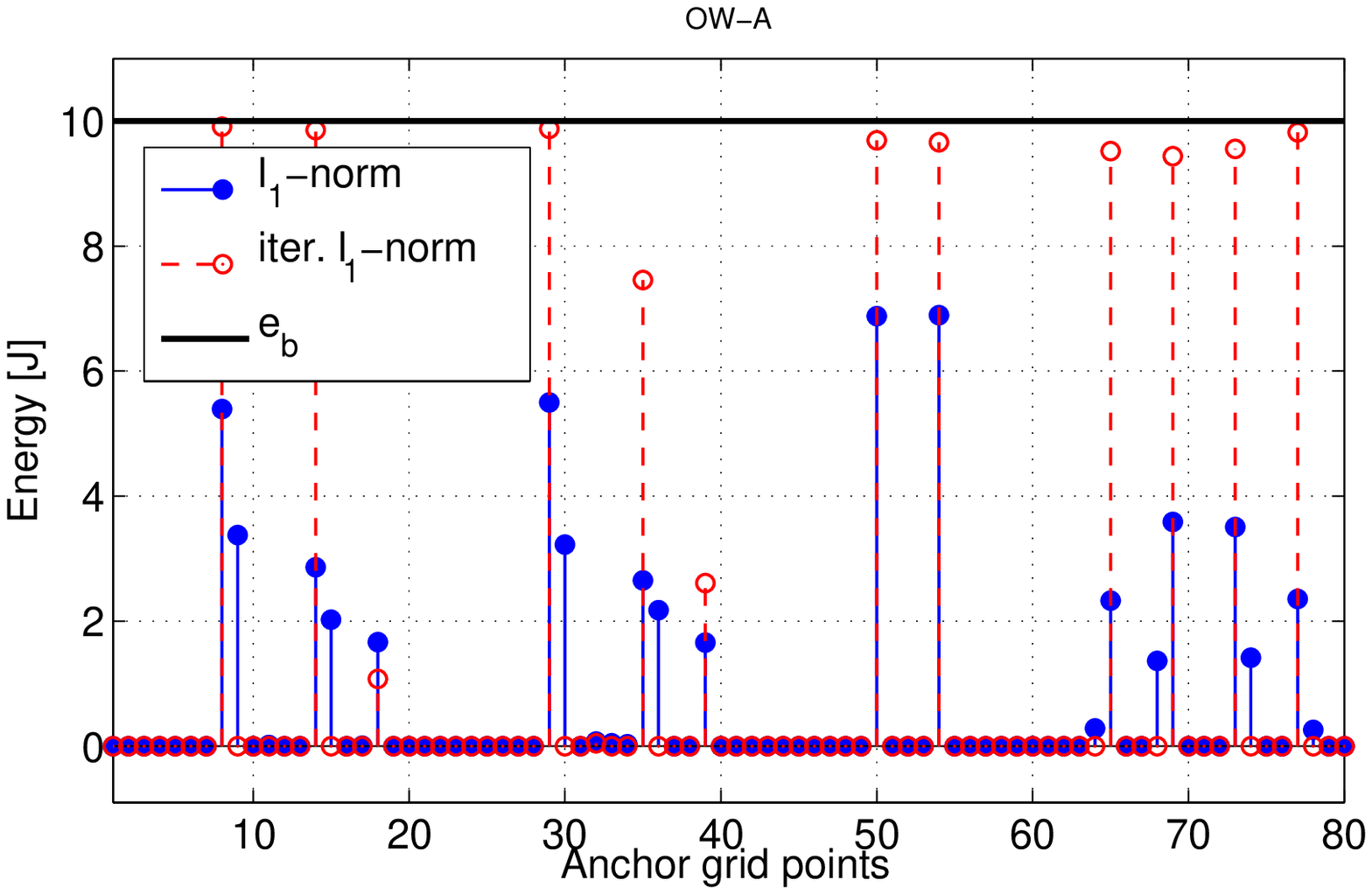}
                \caption{\small Energy allocation.}
                \label{fig:energyowa}
        \end{subfigure}%
        ~ 
                  \begin{subfigure}[t]{0.33\textwidth}
                \centering
                \includegraphics[width=\columnwidth,height=1.75in]{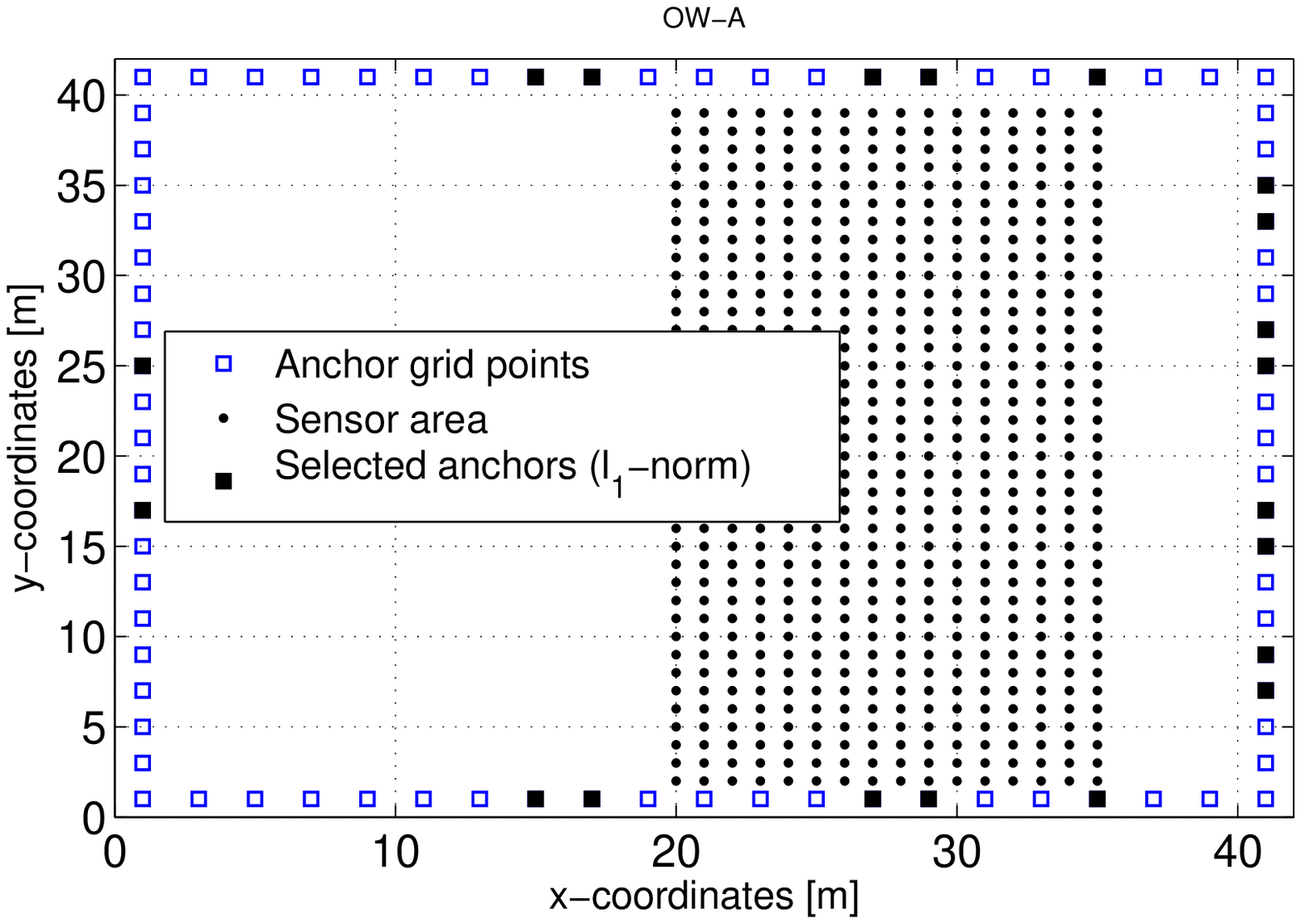}
                \caption{\small $\ell_1$-norm.}
                \label{fig:l1owa}
        \end{subfigure}%
                ~ 
            \begin{subfigure}[t]{0.33\textwidth}
                \centering
                \includegraphics[width=\columnwidth,height=1.75in]{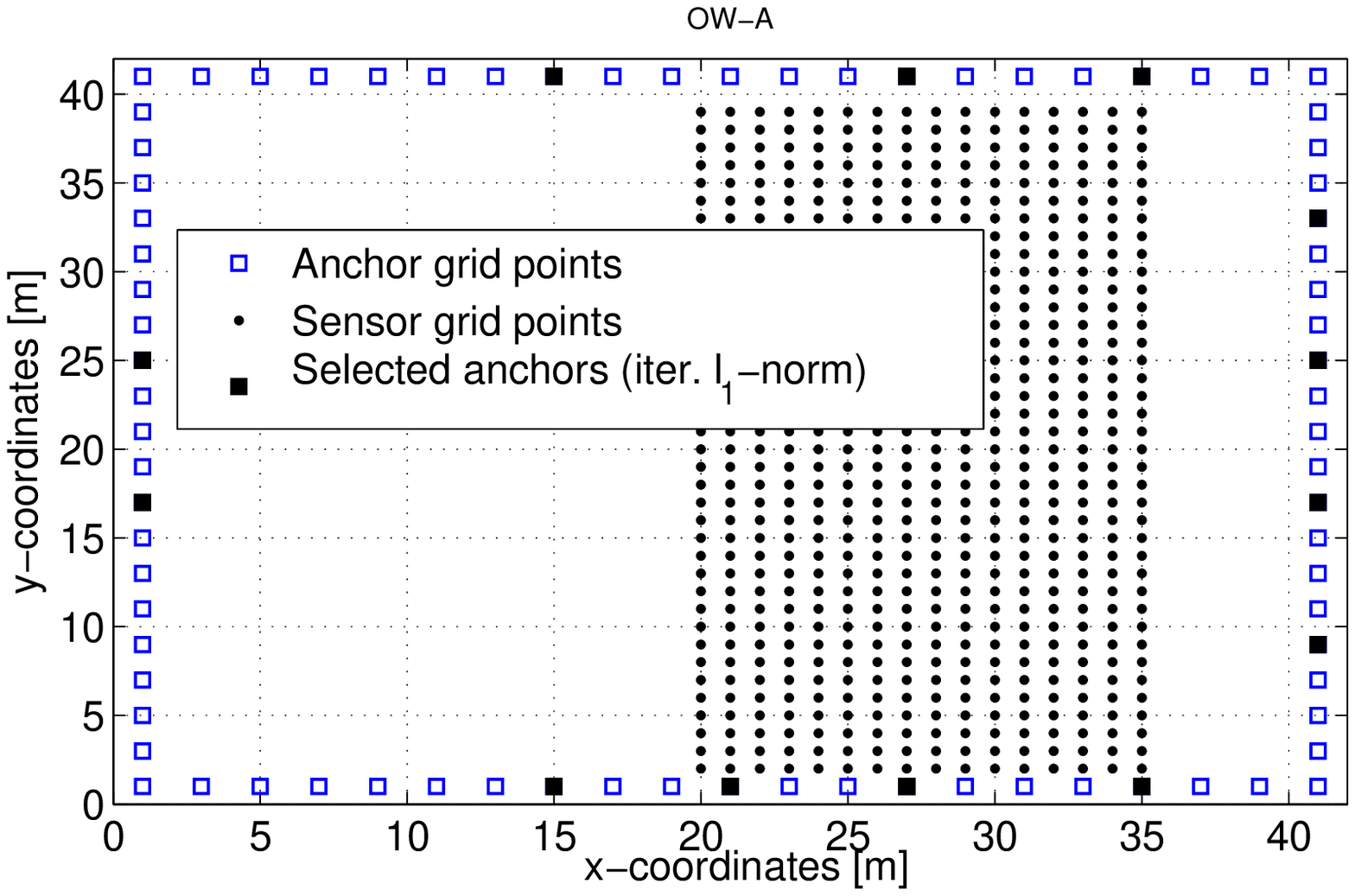}
                 \caption{\small {\it iterative} $\ell_1$-norm.}
                \label{fig:iterl1owa}
        \end{subfigure}  \vspace*{-3mm} 
         \caption{\small Joint anchor placement and energy optimization for OW-A, for $P_e = 0.95$, $e_b=10\mathrm{J}$, and $R_e = 4 \mathrm{cm}$.}\label{fig:OW-A}   \vspace*{-3mm}
\end{figure*}  
\vspace*{-3mm}
\section{Sparsity-exploiting anchor placement}\label{sec:optimization}
\vspace*{-2mm}
Let us now assume that $M$ possible anchors are placed on a discrete grid obtained by uniformly sampling an anchor area $\mathcal{A}$. And remember that the location of the sensor is unknown within a sensor area $\mathcal{S}$. In many localization applications, it makes sense to assume that a limited number of anchors service a prescribed geographical area. This assumption naturally leads to the design of a sparse vector for optimal anchor placement. 
More specifically, we aim to design a sparse {\it joint selection and ranging energy vector} ${\bf e}$ and a sparse {\it selection vector} ${\bf w}$ for OW-A and OW-S, respectively.
\vspace*{-2mm}
\subsection{OW-A: Joint anchor placement and ranging energy optimization}
When the anchors send the ranging signals, the optimization problem can be written as
\begin{equation}
\small
\boxed{
\begin{aligned} \label{eq:opt1}
\min_{{\bf e}\, \in \, \mathbb{R}^M} \quad & {\|{\bf e}\|}_0\\
s.t.  \quad &\sum_{m=1}^{M} e_m {\bf F}_{a,m}({\bf s}) - \lambda {\bf I}_2 \succeq 0, \quad \forall {\bf s} \, \in \, \mathcal{S}\\
&{\bf 0}_{M \times1} \leq {\bf e} \leq e_b{\bf 1}_{M \times1},
\end{aligned}}
\end{equation}
where the $\ell_0$-(quasi) norm refers to the number of non-zero entries of ${\bf e}$, i.e., ${\|{\bf e}\|}_0 := |\{m\,: \, e_m \neq 0\}|$. In addition to the performance constraint, an energy source is positive-valued, and is generally constrained by a prescribed value $e_b$ due to the practical limitations of a source. Here, the threshold $\lambda$ is indirectly the sparsity-inducing parameter (for a fixed $e_b$), where $\lambda \rightarrow 0$ implies a sparser solution. 

It is well known that the $\ell_0$-norm optimization is NP-hard and non-convex. A computationally tractable solution is to use the traditional convex surrogate for the $\ell_0$-norm, which leads to the following optimization problem:
\begin{equation}
\boxed{
\begin{aligned} \label{eq:opt2}
{\bf e}^\ast =& \quad \arg \min_{{\bf e}\, \in \, \mathbb{R}^M} \quad {\bf 1}^T {\bf e}={\|{\bf e}\|}_1\\
s.t.   &\quad \sum_{m=1}^{M} e_m {\bf F}_{a,m}({\bf s}) - \lambda {\bf I}_2 \succeq 0, \quad \forall {\bf s} \, \in \, \mathcal{S}\\
&\quad {\bf 0}_{M \times1} \leq {\bf e} \leq e_b{\bf 1}_{M \times1}.
\end{aligned}}
\end{equation}
The purpose of the $\ell_1$-norm in \eqref{eq:opt2} is twofold: promote the sparsity of the spatially distributed anchors and more importantly minimize the ``total" ranging energy of the network. This is a SDP problem in a standard dual form, and its solution upper bounds the dual feasible ${\bf e}$, i.e., ${\bf e}^\ast \leq {\bf 1}^T{\bf e}$.


The optimization problem in~\eqref{eq:opt2} has multiple global solutions, due to the fact that the $\ell_1$-norm is not strictly convex. 
Even though the $\ell_1$-norm minimizes the total ranging energy of the network, it does not necessarily minimize the number of anchor nodes. On the other hand, the non-convex (intractable) $\ell_0$-norm optimization leads to a higher energy. To improve upon the $\ell_1$-norm solution, we propose an alternative (convex) optimization algorithm, which also results in a correct solution, but with a fewer number of anchors.    
For this purpose, we modify \eqref{eq:opt2} and use the sparsity-enhancing iterative re-weighted $\ell_1$-norm minimization~\cite{candes08} originally used for linear $\ell_1$-regularized least-squares (LS) problems in compressed sensing (CS).
\vspace*{-2mm}
\subsection{Sparsity-enhancing iterative algorithm for OW-A}\label{sec:iter1}
The iterative re-weighted $\ell_1$-norm minimization~\cite{candes08} algorithm is adapted to suit our problem. Let ${\bf u} = [u_1,\ldots,u_M]^T \in \mathbb{R}^{M}$ denote the weight vector.
 The iterative algorithm proceeds as follows:
\begin{enumerate}
\small
\item {\it Initialize} the iteration counter $k=0$ and the weight vector ${\bf u}^{(0)} = {\bf 1}_M$. \vspace*{-3mm}
\item {\it Solve} the weighted $\ell_1$-norm minimization problem
\begin{equation}
\boxed{
\small
\begin{aligned} \label{eq:opt3}
&\min_{{\bf e}^{(k)}\, \in \, \mathbb{R}^M} \quad {\bf u}^{(k)T}{\bf e}^{(k)}\\
&s.t.  \quad \sum_{m=1}^{M} e_m^{(k)} {\bf F}_{a,m}({\bf s}) - \lambda {\bf I}_2 \succeq 0, \quad \forall {\bf s} \, \in \, \mathcal{S}\\
& \quad \quad {\bf 0}_{M \times1} \leq {\bf e}^{(k)} \leq e_b{\bf 1}_{M \times1}.
\end{aligned}}
\end{equation}
for the optimum ${\bf e}^{(k)}$ in the $k$-th iteration.\vspace*{-3mm}
\item {\it Update} the weight vector $u_i^{(k)} = \frac{1}{\epsilon + |e_i^{(k)}|}$, for each $i=1,\ldots,M$.\vspace*{-3mm}
\item {\it Stop} on convergence, or when $k$ attains a specified maximum number of iterations $k_{max}$, else, increment $k$ and go to step~2.
\end{enumerate}
The weight updates force the small entries of the vector ${\bf e}^{(k)}$ to zero and avoid inappropriate suppression of larger entries. The parameter $\epsilon > 0$ is the threshold which provides stability, and guarantees that the zero valued entry of $|{\bf e}^{(k)}|$ does not strictly prohibit a nonzero estimate at the next step. 
\vspace*{-3mm}
\subsection{OW-S: Anchor grid points selection}
In the OW-S case, where the sensor sends the ranging signals, a similar optimization problem can be formulated for the selection of the anchor grid points, which is given by
\begin{equation}
\boxed{
\begin{aligned} \label{eq:opt4}
{\bf w}_{bp}^\ast =&\quad  \arg \min_{{\bf w}  \in \, \{0,1\}^M} \quad {\bf 1}^T{\bf w}={\|{\bf w}\|}_1 \\
s.t. & \quad \sum_{m=1}^{M} w_m {\bf F}_{s,m}({\bf s}) - \lambda {\bf I}_2 \succeq 0, \quad \forall {\bf s} \, \in \, \mathcal{S}.
\end{aligned}}
\end{equation}
The optimization problem in~\eqref{eq:opt4} is a non-convex {\it Boolean} programming problem. However, this can be brought to the standard dual SDP form using the Lagrangian relaxation $w_m(w_m-1)=0$, and introducing a new variable ${\bf W}= {\bf w}{\bf w}^T$ with elements $[{\bf W}]_{mn},m,n=1,\ldots,M$~\cite{Boyd}. The optimization problem is then given by
\begin{equation}
\small
\boxed{
\begin{aligned} \label{eq:opt5}
({\bf w}_{sdp}^\ast, {\bf W}_{sdp}^\ast) =& \arg \min_{{\bf w} \in \mathbb{R}^M, {\bf W} \in  \mathbb{R}^{M \times M}}   {\bf 1}^T{\bf w}={\|{\bf w}\|}_1\\
s.t.  \quad&\sum_{m=1}^{M} w_m {\bf F}_{s,m}({\bf s})- \lambda {\bf I}_2 \succeq 0, \quad \forall {\bf s} \, \in \, \mathcal{S}\\
       \quad&\, {[{\bf W}]}_{mm} = {w}_m, \quad m=1,\ldots,M,\\
      \quad&\left[\begin{array}{cc}{\bf W} & {\bf w} \\{\bf w}^T & 1\end{array}\right] \succeq 0. 
\end{aligned}}
\end{equation}
Here, the rank-$1$ constraint on ${\bf W}$ is relaxed, and \eqref{eq:opt5} can be solved efficiently in polynomial time. The SDP in~\eqref{eq:opt5} provides a good approximation for the  Boolean problem in~\eqref{eq:opt4}, and the solutions for~\eqref{eq:opt4} and~\eqref{eq:opt5} are upper bounds for the dual feasible ${\bf w}$, i.e., ${\bf w}_{bp}^\ast \le {\bf w}_{sdp}^\ast \le {\bf 1}^T{\bf w}$. From ${\bf w}^\ast_{sdp}$, the approximate Boolean solution to ${\bf w}$ can be obtained using randomization techniques.
\vspace*{-3mm}
\subsection{Sparsity-enhancing iterative algorithm for OW-S}\label{sec:iter2}
\vspace*{-2mm}
The sparsity-enhancing iterative algorithm for OW-S can be derived following similar lines as discussed in Section~\ref{sec:iter1}, and it proceeds as follows:
\begin{enumerate}
\small
\item {\it Initialize} the iteration counter $k=0$ and the weight vector ${\bf u}^{(0)} = {\bf 1}_M$. \vspace*{-3mm}
\item {\it Solve} the weighted $\ell_1$-norm minimization problem
\begin{equation}
\small
\boxed{
\begin{aligned} \label{eq:opt6}
&\min_{{\bf w} \in \mathbb{R}^M, {\bf W} \in  \mathbb{R}^{M \times M}} \quad  {\bf u}^{(k)T}{\bf w}\\
s.t.  \quad &\quad \sum_{m=1}^{M} w_m {\bf F}_{s,m}({\bf s})- \lambda {\bf I}_2 \succeq 0, \quad \forall {\bf s} \, \in \, \mathcal{S}\\
       &\quad  {[{\bf W}]}_{mm} = {w}_m, \quad m=1,\ldots,M,\\
      &\quad  \left[\begin{array}{cc}{\bf W} & {\bf w} \\{\bf w}^T & 1\end{array}\right] \succeq 0. 
\end{aligned}}
\end{equation}
for the optimum ${\bf w}^{(k)}$ in the $k$-th iteration.\vspace*{-3mm}
\item {\it Update} the weight vector $u_i^{(k)} = \frac{1}{\epsilon + |w_i^{(k)}|}$, for each $i=1,\ldots,M$.\vspace*{-4mm}
\item {\it Stop} on convergence, or when $k$ attains a specified maximum number of iterations $k_{max}$, else, increment $k$ and go to step~2.
\end{enumerate}
The Boolean solution can again be obtained using randomization.

Note that the sensor ranging energy $e_s$ is not optimized. Once the optimal anchor grid points are obtained, then $e_s$ can be easily optimized on the reduced size problem.    
\begin{figure*}[!t]
\centering
        \begin{subfigure}[t]{0.4\textwidth}
                \centering
                \includegraphics[width=\columnwidth,height=2in]{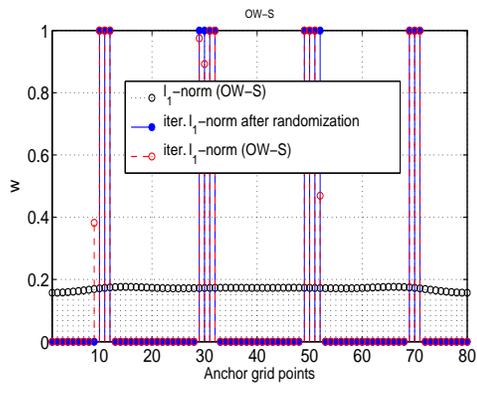} 
\caption{\small $\ell_1$-norm (not randomized) and {\it iterative} $\ell_1$-norm (randomized).}
                \label{fig:selectionows} \vspace*{-3mm}
 \end{subfigure}%
 ~  \begin{subfigure}[t]{0.4\textwidth}
                \centering
                \includegraphics[width=\columnwidth,height=2in]{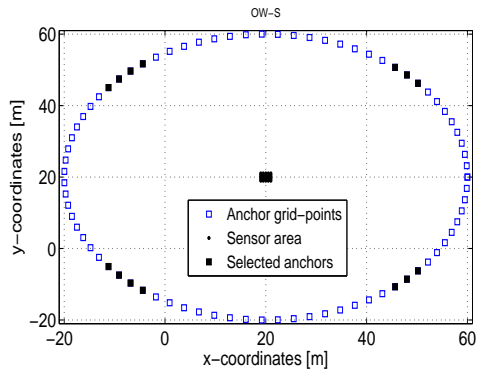}
                \caption{\small Randomized {\it iterative} $\ell_1$-norm solution.}
                \label{fig:iterl1ows}\vspace*{-4mm}
 \end{subfigure}\vspace*{-3mm}
       \caption{\small Anchor placement for OW-S, for $P_e = 0.95$, $e_s=10\mathrm{J}$, and $R_e = 5 \mathrm{cm}$.}\label{fig:OW-S}   \vspace*{-3mm}
\end{figure*}  
\vspace*{-3mm}\section{Simulation results}\label{sec:sim} \vspace*{-3mm}
To test the proposed SDP-based algorithms, we use CVX~\cite{cvx}. CVX in turn calls SeDuMi, a MATLAB implementation of the second-order interior-point methods for computations.

We simulate a network with $M=80$ anchor grid points and for two scenarios of the sensor area $\mathcal{S}$ consisting of 608 and 25 sensor grid points as shown in Fig.~\ref{fig:OW-A} and Fig.~\ref{fig:OW-S}, respectively. Here, we have sampled $\mathcal{S}$ to generate discrete sensor grid points\footnote{A more profound treatment of the gridding of the sensor area can be found in~\cite{Tao09TSP}.}.
We use the following parameters for the simulations: $\alpha=1$, $\beta=2$, $c= 3 \times 10^8 \,\mathrm{ms}^{-1}$, $\frac{\omega}{2\pi} = 8 \, \mathrm{GHz}$, $N_s/2 = 0 \, \mathrm{dBW/Hz}$, and $\epsilon = 10^{-8}$. The constraint on the anchor ranging energy is $e_b = 10 \mathrm{J}$, and the sensor ranging energy $e_s = 10 \mathrm{J}$ is used.

The optimal anchor placement for the OW-A case, based on $\ell_1$-norm minimization of \eqref{eq:opt2} and the {\it iterative} $\ell_1$-norm minimization is illustrated in Fig.~\ref{fig:l1owa} and Fig.~\ref{fig:iterl1owa}, respectively. The algorithm selects the anchor grid points close to the sensor area due to the assumed path-loss model. There exits many similar solutions, however, we are interested in the best solution. This effect can be seen in the not so sparse $\ell_1$-norm solution, while the {\it iterative} $\ell_1$-norm solution enhances the sparsity, yet yielding a correct solution. Fig.~\ref{fig:energyowa} shows the optimal ranging energies that the anchors should adopt.

Fig.~\ref{fig:selectionows} illustrates the anchor selection for a circular anchor grid with $M=80$ points for OW-S. Here, we consider a relatively small sensor area. For the $\ell_1$-norm solution, all the grid points are nominally the same, and the selection weights are spread over all the nodes. However, the {\it iterative} $\ell_1$-norm solution enhances the sparsity. Fig.~\ref{fig:selectionows} also shows an approximate Boolean solution computed using the solution from the {\it iterative} algorithm followed by randomization. Note that the randomization using the $\ell_1$-norm solution will not yield a meaningful Boolean solution, as the $\ell_1$-norm solution is not sparse for this scenario. Fig.~\ref{fig:iterl1ows} shows the corresponding anchor placement using the {\it iterative} $\ell_1$-norm algorithm followed by randomization. 

An exhaustive search based algorithm to select 14 grid points out of 80 for the case shown in  Fig.~\ref{fig:iterl1owa} would need $\binom{80}{13} \times 608 \approx 10^{17}$ searches over the constraint set, which is clearly intractable. On the other hand, the complexity of the proposed algorithms for both OW-A and OW-S cases are polynomial in the number of variables and the number of constraints.
\vspace*{-3mm}
\bibliographystyle{IEEEbib}
\bibliography{IEEEabrv,strings,refs}


\end{document}